\documentclass[a4paper,12pt, epsfig]{article}
\usepackage{epsfig}
\usepackage{amssymb}
\usepackage{amsfonts}
\usepackage{amsmath}
\usepackage{subfig}
\usepackage{wrapfig}

\newskip\humongous \humongous=0pt plus 1000pt minus 1000pt

\newif\ifdtup

\catcode`\@=11

\@addtoreset{equation}{section}
\def\theequation{\arabic{equation}}

\def\@normalsize{\@setsize\normalsize{15pt}\xiipt\@xiipt
\abovedisplayskip 14pt plus3pt minus3pt%
\belowdisplayskip \abovedisplayskip
\abovedisplayshortskip \z@ plus3pt%
\belowdisplayshortskip 7pt plus3.5pt minus0pt}

\def\small{\@setsize\small{13.6pt}\xipt\@xipt
\abovedisplayskip 13pt plus3pt minus3pt%
\belowdisplayskip \abovedisplayskip
\abovedisplayshortskip \z@ plus3pt%
\belowdisplayshortskip 7pt plus3.5pt minus0pt
\def\@listi{\parsep 4.5pt plus 2pt minus 1pt
     \itemsep \parsep
     \topsep 9pt plus 3pt minus 3pt}}

\relax

\catcode`@=12

\catcode`\@=11

\def\section{\@startsection{section}{1}{\z@}{3.5ex plus 1ex minus
   .2ex}{2.3ex plus .2ex}{\large\bf}}

\def\thesection{\arabic{section}}
\def\thesubsection{\arabic{section}.\arabic{subsection}}

\def\appendix{\setcounter{section}{0}
 \def\thesection{Appendix \Alph{section}}
 \def\thesubsection{\Alph{section}.\arabic{subsection}}
 \def\theequation{\arabic{equation}}}

\def\SymBoxes#1#2#3#4{\newdimen\un@t \un@t#3%
\raisebox{#1}{\rule{#2\un@t}{#4}\hskip-#2\un@t
\@tempdimb\un@t \advance\@tempdimb by-#4\@tempcntb#2\relax%
\@whilenum{\@tempcntb>0}\do{
\rule{#4}{\un@t}\hskip\@tempdimb \advance\@tempcntb by\m@ne}%
\hskip-#2\un@t \rule[\un@t]{#2\un@t}{#4}%
\rule[\un@t]{#4}{#4}\hskip-#4
\rule{#4}{\un@t}}\hskip-#4}                

\begin{document}

\newcommand{\beq}{\begin{equation}}
\newcommand{\eeq}{\end{equation}}
\newcommand{\bea}{\begin{eqnarray}}
\newcommand{\eea}{\end{eqnarray}}
\newcommand{\beas}{\begin{eqnarray*}}
\newcommand{\eeas}{\end{eqnarray*}}
\newcommand{\defi}{\stackrel{\rm def}{=}}
\newcommand{\non}{\nonumber}
\newcommand{\bquo}{\begin{quote}}
\newcommand{\enqu}{\end{quote}}
\renewcommand{\(}{\begin{equation}}
\renewcommand{\)}{\end{equation}}
\def\IZ{{\mathbb Z}}
\def\IR{{\mathbb R}}
\def\IC{{\mathbb C}}
\def\IQ{{\mathbb Q}}
\def\IP{{\mathbb P}}

\def \eqn#1#2{\begin{equation}#2\label{#1}\end{equation}}
\def\de{\partial}
\def\Tr{ \hbox{\rm Tr}}
\def\H{ \hbox{\rm H}}
\def\HE{ \hbox{$\rm H^{even}$}}
\def\HO{ \hbox{$\rm H^{odd}$}}
\def\K{ \hbox{\rm K}}
\def\Im{ \hbox{\rm Im}}
\def\Ker{ \hbox{\rm Ker}}
\def\const{\hbox {\rm const.}}
\def\o{\over}
\def\im{\hbox{\rm Im}}
\def\re{\hbox{\rm Re}}
\def\bra{\langle}\def\ket{\rangle}
\def\Arg{\hbox {\rm Arg}}
\def\Re{\hbox {\rm Re}}
\def\Im{\hbox {\rm Im}}
\def\exo{\hbox {\rm exp}}
\def\diag{\hbox{\rm diag}}
\def\longvert{{\rule[-2mm]{0.1mm}{7mm}}\,}
\def\a{\alpha}
\def\dag{{}^{\dagger}}
\def\tq{{\widetilde q}}
\def\p{{}^{\prime}}
\def\W{W}
\def\N{{\cal N}}
\def\hsp{,\hspace{.7cm}}
\newcommand{\C}{\ensuremath{\mathbb C}}
\newcommand{\Z}{\ensuremath{\mathbb Z}}
\newcommand{\R}{\ensuremath{\mathbb R}}
\newcommand{\rp}{\ensuremath{\mathbb {RP}}}
\newcommand{\cp}{\ensuremath{\mathbb {CP}}}
\newcommand{\vac}{\ensuremath{|0\rangle}}
\newcommand{\vact}{\ensuremath{|00\rangle}}
\newcommand{\oc}{\ensuremath{\overline{c}}}
\begin{titlepage}
\begin{flushright}
\end{flushright}
\bigskip
\def\thefootnote{\fnsymbol{footnote}}

\begin{center}
{\Large {\bf A Comment on Kerr-CFT and Wald Entropy \\
}}
\end{center}

\bigskip
\begin{center}
{\large  Chethan
KRISHNAN$^1$\footnote{\texttt{Chethan.Krishnan@ulb.ac.be}} \ and \ Stanislav KUPERSTEIN$^2$ \footnote{\texttt{skuperst@vub.ac.be }}}
\end{center}

\renewcommand{\thefootnote}{\arabic{footnote}}

\begin{center}
\vspace{1em}
{\em  { $^1$International Solvay Institutes,\\
Physique Th\'eorique et Math\'ematique,\\
ULB C.P. 231, Universit\'e Libre
de Bruxelles, \\ B-1050, Bruxelles, Belgium\\}}

\vspace{1em}
{\em  { $^2$Theoretische Natuurkunde,\\
Vrije Universiteit Brussel and The International Solvay Institutes\\
Pleinlaan 2, B-1050 Brussels, Belgium \\}}

\end{center}

\noindent
\begin{center} {\bf Abstract} \end{center}
We point out that the entropies of black holes in general diffeomorphism invariant theories, computed using the Kerr-CFT correspondence and the Wald formula (as implemented in the entropy function formalism), need not always agree. A simple way to illustrate this is to consider Einstein-Gauss-Bonnet gravity in four dimensions, where the Gauss-Bonnet term is topological. This means that the central charge of Kerr-CFT computed in the Barnich-Brandt-Compere formalism remains the same as in Einstein gravity, while the entropy computed using the entropy function gives a universal correction proportional to the Gauss-Bonnet coupling. We argue that at least in this example, the Kerr-CFT result is the physically reasonable one.  The resolution to this discrepancy might lie in a better understanding of boundary terms. 

\vspace{1.6 cm}
\begin{center}
\end{center}
\vfill

\end{titlepage}
\bigskip

\hfill{}
\bigskip


\setcounter{footnote}{0}
\section*{\bf Introduction} \label{intro}

\noindent
The Kerr-CFT correspondence is the observation that the entropy of a large class of extremal rotating black holes can be computed using the central charge of a certain two-dimensional conformal algebra that arises in the near horizon geometry of the black hole \cite{KerrCFT, KerrCFT1}.

We demonstrate the basic idea of the correspondence by starting with an extremal black hole metric and taking its near horizon limit. It turns out that extremal black holes in four dimensions generically \cite{Kunduri} have a near horizon geometry (NHG) that takes the form 
\bea\label{NHG}
ds^2=\Gamma(\theta) \left[-r^2dt^2 +\frac{dr^2}{r^2}+\alpha(\theta)d\theta^2\right]+\gamma(\theta)(d\phi+kr dt)^2.
\eea
This geometry is fixed by its isometry group: $SO(2,1) \times U(1)$.
The results of \cite{KerrCFT, KerrCFT1} showed that the charges that generate the asymptotic symmetries \cite{Glenn} of this geometry form a Virasoro algebra with central charge
\bea\label{cL}
c_{L}=3k \int_0^{\pi} d\theta \sqrt{\Gamma(\theta)\alpha(\theta)\gamma(\theta)}.
\eea
This result leads to the natural conjecture that quantum gravity in the background defined by the near-horizon metric is dual to a CFT whose central charge is the one presented above. One can also associate a temperature to the near horizon geometry (we will discuss this in more detail later) by applying the laws of black hole thermodynamics to the black hole spacetime, and the result turns out to be
\bea\label{T}
T_L=\frac{1}{2\pi k}.
\eea
The entropy of the CFT then can be found using a version of the Cardy formula
\bea
\label{KC-entropy}
S=\frac{\pi^2}{3}c_L T_L=\frac{\pi}{2}\int_0^{\pi} d\theta \sqrt{\Gamma(\theta)\alpha(\theta)\gamma(\theta)}.
\eea
This coincides with the Bekenstein-Hawking area/entropy of the black hole we started with, which can be computed for generic extremal black holes with the NHG given by (\ref{NHG}) using the Wald formula \cite{Wald1,Wald2} in Sen's entropy function formalism \cite{Sen1,Sen2,Sen3}. Even though the microscopic description of the CFT degrees of freedom are not known here (unlike in the case of the famous Strominger-Vafa black hole \cite{StromingerVafa} and its descendants), the presence of the Virasoro algebra with the correct central charge is a remarkable piece of evidence that even ordinary (extremal) Kerr black holes might have a dual unitary description.

One of the main results of \cite{KerrCFT, KerrCFT1} and followup papers \cite{followers, Pope, Pope1, Geoffrey} is the application of the formula for the central charge for Einstein-Hilbert gravity, computed in \cite{Glenn} (see also \cite{Glenn1, geoffrey}.) to the near-horizon geometries arising from a large array of extreme black holes. In all of these cases the CFT entropy matched the Bekenstein-Hawking entropy. But, as observed in \cite{KerrCFT1, Pope1} and pointed out above, these geometries have a standard form and so it is not clear, to what extent these computations can be considered as distinct checks of the Kerr-CFT proposal. One way to consider a more general situation is to consider a higher derivative theory of gravity, where the central charge formula and the Wald entropy both can potentially change its form, and to see whether the central charge can still reproduce the geometric entropy. This is the context of this note.

We start by noticing that an interesting simplification happens when one looks at Gauss-Bonnet corrected gravity. The advantage of the Gauss-Bonnet term is that it retains the two-derivative nature of gravity, and therefore does not require additions to the {\em formalism} of \cite{Glenn} even though the final central charge does change because the action is different\footnote{More general higher derivative terms typically give rise to higher order equations of motion and would require a modification of, for example, eqn. (1.13) in \cite{Glenn}.}. An even bigger simplification happens, if we look at Gauss-Bonnet corrected gravity in four dimensions, because in this case the Gauss-Bonnet term is a total derivative and does not affect the equations of motion, and this results in no corrections to the central charge expression at all.

This gives us a simple way of exploring the relation between the Kerr-CFT entropy and the Wald entropy. Since the central charge is unchanged, we know that the Kerr-CFT entropy associated with (\ref{NHG}) is the same as in Einstein gravity\footnote{This also involves the assumption that the Frolov-Thorne temperature associated with the background is unchanged in the new context. We present some arguments why we believe this is justified, in section 4.} and is captured in (\ref{KC-entropy}). On the other hand, we can also try to compute the Wald entropy associated to the NHG in Einstein-Gauss-Bonnet theory using Sen's entropy function formalism \cite{Sen2,Sen3}. 
This can be accomplished through a minor extension of the results in \cite{Sen1}. When we do this, we find that the black hole entropy gets a correction proportional to the Gauss-Bonnet coupling, resulting in a mismatch with (\ref{KC-entropy}).

Our result is easily extended to the case when the extremal (rotating) black hole we consider is charged, and is coupled to scalars and Abelian vectors. This is accomplished straightforwardly through the attractor formalism\footnote{A cosmological constant can also be added since its effects are implicit in the near-horizon geometry.} of \cite{Sen1, Sen3}. As long as the Gauss-Bonnet term is topological, the correction in entropy is universal. To keep the essential idea from drowning in the details, we will only consider black holes with rotation here. Some aspects of Kerr-CFT in the context of attractors have been considered in \cite{Astefanesei}. Gauss-Bonnet terms in AdS have been studied in \cite{Olea}, where some related observations for non-rotating black holes have also been made.


\section*{Einstein-Gauss-Bonnet Gravity}

The theory we will consider is Einstein gravity with a Gauss-Bonnet correction in four dimensions:
\bea
\label{action}
&&S=
\int d^4 x \sqrt{-g} \ {\cal L}\equiv\int d^4 x \sqrt{-g} \ ({\cal L}_E+ {\cal L}_{GB})= \int d^4 x \sqrt{-g}\left(R+\xi B\right), \nonumber \\
&&\ \hspace{1.25in}{\rm where} \ \ B=R^2-4 R_{\mu\nu}R^{\mu\nu}+ R_{\alpha\beta\gamma\delta}R^{\alpha\beta\gamma\delta}.
\eea
Here $\xi$ is the Gauss-Bonnet coupling and we have set $16 \pi G=1$. In four dimensions, the integral of $\sqrt{-g} \ B$ is topological. Therefore the equations of motion, which are detremined by local physics, are unchanged from the Einstein equations \cite{Lanczos}. The formalism of asymptotic central charges developed in \cite{Glenn} and used in the Kerr-CFT correspondence depends only on the equations of motion, and therefore we immediately come to the conclusion that the central charges in four-dimensional Einstein-Gauss-Bonnet theory are identical to that in standard Einstein gravity. The explicit form of this central charge can be found in \cite{Glenn, KerrCFT}, but we will only need its specific form for the NHG metric (\ref{NHG}). We present it in (\ref{cL}), it was first computed in \cite{KerrCFT1}.

\section*{Entropy Function}

Sen's entropy function formalism depends on the Lagrangian directly, and not merely on the equations of motion. We will presently compute the entropy associated with (\ref{NHG}) in Gauss-Bonnet gravity using the entropy function. The technology required is presented in \cite{Sen1} for Einstein gravity, and fortunately, we need only minor modifications to adapt it to include Gauss-Bonnet\footnote{It turns out that this result was already noted, before Kerr-CFT, in \cite{Cho}. We apologize to the authors for the omission, and thank Yuji Tachikawa for bringing the paper to our attention.}. This is again (as it turns out) because the Gauss-Bonnet term is topological in 4 dimensions.

The entropy function formalism for the rotating extremal black hole starts out by assuming that the near-horizon geometry has an $SO(2,1) \times U(1)$ isometry as in (\ref{NHG}). On top of this condition, it also assumes that angular part of the near horizon geometry is a smooth deformation of the sphere \cite{Sen3}. This latter condition is a very weak one and is often (if not always) satisfied in the examples considered in the Kerr-CFT context. This means, in particular that the $\theta$ variable in (\ref{NHG}) can be redefined in a more convenient way as in \cite{Sen1, Sen3}:
\bea \label{NHGsen}
&& ds^2 = \Omega(\theta)^2 e^{2\psi(\theta)} (- r^2 dt^2 + dr^2 / r^2 +
\beta^2 d\theta^2) + e^{-2\psi(\theta)} (d\phi+k \ r dt)^2,
\eea
where $\beta$ is a constant. To ensure regularity of the solution, we also need to impose that
\bea \label{reg1}
\Omega(\theta) e^{\psi(\theta)} \to \hbox{constant as
$\theta\to 0,\pi$}\, , \nonumber\\
\beta \Omega(\theta) e^{2\psi(\theta)} \sin\theta \to 1\quad \hbox{
as $\theta\to 0, \pi$}\, .
\eea
Just from the regularity of the background, we therefore end up with
\bea \label{reg2}
&& \Omega(\theta)\to a_0\sin\theta, \quad e^{\psi(\theta)}\to
{1\over \sqrt{\beta a_0} \sin\theta}, \quad \hbox{as $\theta\to 0$},
\nonumber \\
&& \Omega(\theta)\to a_\pi\sin\theta, \quad e^{\psi(\theta)}\to
{1\over \sqrt{\beta a_\pi} \sin\theta}, \quad
\hbox{as $\theta\to \pi$}\, ,
\eea
where $a_0$ and $a_\pi$ are arbitrary constants at this stage.

The entropy function in our case is defined as
\bea \label{EF}
{\cal E}\equiv 2 \pi\left(J k-\int d\theta d\phi \sqrt{-g} \ ({\cal L}_E+ {\cal L}_{GB} ) \right) \hspace{+1.2in} \nonumber \\
=2\pi J k +4\pi^2\,   \left[2 \Omega(\theta)\beta^{-1}
(\psi'(\theta) + 2\Omega'(\theta) / \Omega
(\theta) )\right]_{\theta=0}^{\theta=\pi} -4\pi^2\int d\theta  \sqrt{-g}\ {\cal L}_{GB} + \hspace{0.2in} \nonumber \\
-4\pi^2 \int d\theta \, \Bigg[2\Omega(\theta)^{-1} \beta^{-1}
(\Omega'(\theta))^2 -
2\Omega(\theta) \beta - 2 \Omega(\theta) \beta^{-1}
(\psi'(\theta))^2 +{1\over 2}
\alpha^2 \Omega(\theta)^{-1} \beta e^{-4\psi(\theta)}\Bigg].
\eea
The idea behind Sen's entropy function formalism is that extremizing this expression with respect to the undetermined parameters and functions of the near-horizon geometry ($\Omega(\theta), \psi(\theta)$ and $\beta$) results in the entropy of the black hole. $J$ above will be identified as the angular momentum of the black hole. In the expression above, we have explicitly expanded the Einstein piece, and taken out a total derivative in the Einstein action and expressed it as a  boundary term. As emphasized in \cite{Sen1}, the boundary piece we have written above has the property that it vanishes under boundary variations satisfying (\ref{reg1}, \ref{reg2}), and this guarantees that standard variational principles can be immediately applied without worrying about the constraints arising from (\ref{reg1}, \ref{reg2}). This is a major technical advantage.

But what about the Gauss-Bonnet term? We have not explicitly written its form because it is a complicated mess. But the redeeming fact that saves the day is that fortunately it is a total derivative, and can be re-expressed as a boundary term. The fundamental reason why this is possible is because the Gauss-Bonnet term is topological in four dimensions. It turns out that for the background (\ref{NHGsen}) it can be written as
\bea
\sqrt{-g}\ {\cal L}_{GB}=-\xi\frac{d}{d \theta}\Bigg[\frac{2 e^{-6 \psi (\theta )}   \psi '(\theta ) \left(-5 k^2 \beta ^2+4 e^{4 \psi (\theta )} \Omega (\theta )^2 \left(\beta ^2+\psi '(\theta )^2\right)\right)}{\beta ^3 \Omega (\theta )^3}+\nonumber \\
+\frac{2  \left(-\frac{2 e^{-6 \psi (\theta )} k ^2 \beta ^2 \Omega '(\theta)} {\Omega (\theta )^4}+e^{-2 \psi (\theta )} \left(\frac{8 \psi '(\theta )^2 \Omega '(\theta )}{\Omega (\theta )^2}+\frac{4 \psi '(\theta ) \Omega '(\theta )^2}{\Omega (\theta )^3}\right)\right)}{\beta ^3}\bigg]\equiv -\xi \frac{d}{d \theta} X(\theta).
\eea
The boundary terms arising from integration of this term in (\ref{EF}) can easily be checked to vanish\footnote{Note that since $\beta$ is a parameter and not a function, its variations have to be identical at the boundary and in the bulk, and therefore is taken care of by the compensating variations of the fields satisfying the bulk equations of motion.} under boundary variations satisfying the regularity conditions (\ref{reg1}, \ref{reg2}). Therefore our bulk equations of motion reduce to those found in \cite{Sen1} (on setting the scalars  and vectors to zero.). In our particular case, where we are dealing with pure gravity, these can be solved explicitly, and we end up with rather simple forms for $\Omega(\theta), \  \psi(\theta)$ and $\beta$:
\bea
\Omega(\theta)=a \sin \theta, \ \ \  e^{\psi(\theta)}=\frac{1}{2} \sqrt{\frac{\cot\left(\frac{\theta }{2}\right)^2+k ^2 \tan\left(\frac{\theta }{2}\right)^2}{a}}, \ \
 \beta=1.
\eea
The $a_0$ and $a_\pi$ introduced earlier are both forced to be equal to $a$, an arbitrary constant. If the NHG is thought of as arising from a full black hole solution, it will be determined by the full black hole solution from which the near horizon geometry arises. The symmetry and regularity allow us to fix the near-horizon geometry {\em explicitly} in terms of two parameters, $a$ and $k$. In a full black hole solution, these parameters will of course not be independent because (ignoring the Gauss-Bonnet term for now) we are working with Einstein gravity where the extremal black hole is Kerr and is completely fixed by its angular momentum. It can be checked that the near horizon Kerr metric written down in \cite{KerrCFT} can be translated into the form we have written here. It is also important to remember that in more generic cases, the coordinate transformation that relates an NHG of the form (\ref{NHG}) with the form (\ref{NHGsen}) involves a fixing of the ranges of the $\theta$ coordinate.

With the equations arising from extremizing the near-horizon variables, one can obtain expressions for the entropy and the angular momentum.
\bea
J &=&{4\pi\over k}  \Big[\Omega \psi'\Big]^{\theta=\pi}_{\theta=0} \ , \\
S&=&8 \pi^2\left[-2 \Omega'+{\Omega \over \beta}\left(\psi'+ 2 {\Omega' \over
\Omega}\right)\right]^{\theta=\pi}_{\theta=0}+4\pi^2 \xi \Big[X(\theta)\Big]^{\theta=\pi}_{\theta=0} \label{tot}
\eea
The second term in the last line comes from the Gauss-bonnet coupling. By plugging in the the explicit functions or their boundary forms as given in (\ref{reg1}, \ref{reg2}), we find the angular momentum and the Wald entropy of the extremal Gauss-Bonnet black hole:
\bea
J=\frac{8 \pi a}{k}, \  \  \ S=16 \pi^2 a+ 64 \pi^2 \xi. \label{senen}
\eea
We emphasize that this result does not change if we couple the theory minimally to scalars and vectors, so it holds also for charged black holes. This is because they are determined already by the (\ref{reg1}, \ref{reg2}), and do not depend on our ability to explicitly solve for $\psi(\theta)$. Unlike $\Omega(\theta)$ and $\beta$, solving $\psi(\theta)$ explicitly is easy only in pure gravity. 

\section*{The Frolov-Thorne Temperature}

The Frolov-Thorne temperature $T_L$ \cite{FT} is a quantity that is computed for a given {\em background}, and therefore is independent of the specific action one is working with. Its construction as given in \cite{KerrCFT}, depends on the original extremal black hole background that we started with in order to get to the NHG. In general theories of gravity (including Gauss-Bonnet in dimensions higher than four), such full exact rotating solutions are not known. But in four dimensions, since the equations of motion are unaffected, the extremal Kerr-solution is still a solution. This means that the arguments of section 8 in \cite{KerrCFT} can be applied here, and one gets the same result, $T_L=1/2\pi$. Notice that when we are working with Kerr, we are working specifically with $k=1$, so this is consistent with the general expression (\ref{T}).

Another way to motivate that the Frolov-Thorne temperature is unchanged in our case (from that in Einstein gravity) is to use the first law of Black Hole thermodynamics specialized to the case of extremal black holes, as done in Section 5 of \cite{KerrCFT1}. In our case, the relevant equation takes the form $dS= dJ/T_L$. Note that the expression for $J$ above (and therefore $dJ$) is identical to that in Einstein gravity, while since the correction in entropy is by a constant piece, $dS$ is also exactly as in Einstein gravity. This means that $T_L$ also should not change, and must take the form that it took in Einstein gravity,
\bea
T_L=\frac{1}{2\pi k}.
\eea
Either way, even though a completely unambiguous definition of the Frolov-Thorne temperature in general is not available, we believe that the formula above is sufficiently well-motivated in our context.

\section*{The Central Charge and the Comparison}

The central charge expression (\ref{cL}) for the background (\ref{NHGsen}) takes the form
\bea
c_L=3k\int_0^{\pi}a \sin \theta d\theta =6 ak.
\eea
We emphasize that since the expressions for the charges that generate the asymptotic symmetry group are unchanged here, the fall-offs considered in \cite{KerrCFT,KerrCFT1} will ensure that the charges are finite here as well.
In the end then, using the Cardy formula (\ref{KC-entropy}) and the central charge above, we get the entropy to be given by
\bea
S=\pi a.
\eea
Remembering that we were working with $16 \pi G=1$ in the section on entropy function, we see that this precisely reproduces the Einstein part of the entropy in (\ref{senen}). But the Gauss-Bonnet piece in the entropy function is not reproduced by the Kerr-CFT computation. We discuss some aspects of this result in the next section.

\section*{Comments, Caveats and Speculations}

The point of the simple computation we have done in this paper is as a proof of principle. It demonstrates that with the standard definitions on either side, there is a tension between the Wald formula (as used in the entropy function approach) and the Kerr-CFT correspondence. The entropy correction we find arises (surprisingly) from a total derivative piece in the action, in the resulting entropy shift is a constant for all extremal Gauss-Bonnet black holes, proportional to the Gauss-Bonnet coupling. In the context of black hole thermodynamics, such a shift in the datum would not be significant since only differentials matter, so it might be the case that this ``discrepancy" is not really discrepancy. But even in this case, it is important to understand precisely where the formalism fails. When we take the angular momentum (and therefore mass) of the black hole to zero in the entropy function result (\ref{senen}), we find that the spacetime is left with a finite entropy. This is already an indication that the result there might not be fully trustable\footnote{We thank Amitabh Virmani for this comment.}. More importantly, the absolute entropy of the black hole is taken to directly be a measure of the number of microscopic degrees of freedom and is often the basis of many successful computations in string theory \cite{StromingerVafa}. 

One direction one can pursue in order to try to understand this result better, is to explore higher derivative (in the EOM) theories. But before that, it would be interesting to consider higher dimensional versions of the Gauss-Bonnet case we have discussed here. The equations of motion are still second order, so the Barnich-Brandt-Compere approach can be directly applied to the theory \cite{Glenn} to compute the modified central charge\footnote{The result would in general  differ from the Einstein gravity result of \cite{Glenn, KerrCFT}, of course.}. In dimensions greater than 4 however, the Gauss-Bonnet term is no longer topological, so we will find non-constant corrections to the entropy. This makes it a richer playground for exploring the ideas of this paper. But the entropy function formalism in higher dimensions is more complicated because the Gauss-Bonnet term is no longer topological\footnote{We emphasize that the full determination of all the near horizon functions is not necessary for the determination of the entropy and the charges. 
}. This will perhaps require one to work perturbatively in $\xi$. Some of these questions are currently being investigated \cite{KK}. Similar statements can also be made about black holes in Einstein-Lovelock gravity \cite{Gaston}, which is the most general setting to explore sensible two derivative (pure) gravity theories with higher order curvature corrections. Whenever there are topological terms like in our case (another example is provided by the degree $n$ Lovelock term in $2n$ dimensions), its seems natural that the formalism should be modified to include boundary terms. It would be interesting to construct a formalism that incorporates the effects of York-Gibbons-Hawking like terms \cite{York, Gibbons}.  

From the structure of our result, it is evident that if we drop the total derivative piece in the Lagrangian, the agreement between Wald formula and the Kerr-CFT result is immediately restored. In conventional Einstein gravity also there are total derivative pieces: in fact, this is the reason why we get second order equations from the Einstein action, despite the fact that the Ricci scalar contains second derivatives of the metric. In the entropy formula (\ref{tot}), the contributions from this total derivative piece (which also shows up in (\ref{EF})) adds up to zero. So it is tempting to speculate that the successes of Wald entropy are retained if we declare by fiat that the total derivative pieces in the action be dropped when computing the entropy\footnote{Most tests of Wald formula depend only on pieces that are not total derivatives. For example, the successful checks of microstate counting in Gauss-Bonnet corrected $D=4, {\cal N}=4$ black holes string theory do not deal with topological terms because of non-trivial scalar couplings.}. The question will be to construct a covariant formalism where this can be implemented. A consistent way of taking account of these boundary terms might lie along the lines of \cite{Glenn1, geoffrey, marolf} \footnote{On a different note, we thank Glenn Barnich for emphasizing to us some of the ambiguities in the Wald formula, in particular, regarding asymptotic Killing vectors. Geoffrey Compere informs us that a paper dealing with these ambiguities in the context of Kerr-CFT, is on its way \cite{geoffreynew}.}.

Another point we should mention is that the original Wald formula was derived for bifurcate Killing horizons, and extremal black hole horizons are not bifurcate. But this has never been a problem in the construction and successes of the entropy function formalism, because the entropy of the extremal black hole here is {\em defined} as that of an ordinary black hole, with a bifurcate horizon, as it approaches the limit of extremality \cite{Sen3}.

Finally, it would be most interesting to generalize the Kerr-CFT correspondence to include non-extremal black holes. See \cite{Mu} for related discussions. Understanding the non-extremal case would be directly relevant to real world black holes, as well as to the plethora of vacuum solutions to higher dimensional Einstein gravity constructed in the past years \cite{5d}. 

\section*{Acknowledgments}

It is our pleasure to thank Glenn Barnich for discussions, clarifications and encouragement, Amitabh Virmani for a careful reading of the manuscript and Geoffrey Compere for correspondence. CK would like to thank Glenn Barnich and Amitabh Virmani for a related previous collaboration and the audience at Solvay for questions and comments during a talk on the subject. We would also like to acknowledge the organizers and participants of the workshop on 3D gravity at the Winter School in CERN (9-13 Feb. 2009) for a stimulating environment that triggered this work. This work is supported in part by IISN - Belgium (convention 4.4505.86), by the Belgian National Lottery, by the
European Commission FP6 RTN programme MRTN-CT-2004-005104 in which the
authors are associated with V. U. Brussel, and by the Belgian Federal
Science Policy Office through the Interuniversity Attraction Pole P5/27.




\end{document}